# Next nearest neighbour coupling with spinor polariton condensates

Dmitriy Dovzhenko,[1, *] Denis Aristov,[1] Lucinda Pickup,[1] Helgi Sigurðsson,[2, 3] and Pavlos Lagoudakis[1, 4]

[1]*School of Physics and Astronomy, University of Southampton, Southampton, SO17 1BJ, UK*
[2]*Science Institute, University of Iceland, Dunhagi 3, IS-107, Reykjavik, Iceland*
[3]*Institute of Experimental Physics, Faculty of Physics,*
*University of Warsaw, ul. Pasteura 5, PL-02-093 Warsaw, Poland*
[4]*Hybrid Photonics Laboratory, Skolkovo Institute of Science and Technology,*
*Territory of Innovation Center Skolkovo, Bolshoy Boulevard 30, building 1, 121205 Moscow, Russia*
(Dated: August 2, 2024)

We report on experimental observation of next-nearest-neighbour coupling between ballistically expanding spinor exciton-polariton condensates in a planar semiconductor microcavity. All-optical control over the coupling strength between neighbouring condensates is demonstrated through distance-periodic pseudospin screening of their ballistic particle outflow due to the inherent splitting of the planar cavity transverse-electric (TE) and transverse-magnetic (TM) modes. By screening the nearest-neighbour coupling we overcome the conventional spatial coupling hierarchy between condensates. This offers a promising route towards creating unconventional non-planar many-body Hamiltonians using networks of ballistically expanding spinor exciton-polariton condensates.

Strongly correlated quantum many-body systems have attracted a lot of interest as a promising tool to engineer and explore phases of matter in extreme settings [1–3] and to simulate complex Hamiltonians [4, 5]. Such systems include ultracold atomic ensembles [4], trapped ions [6, 7], nuclear and electronic spins [8, 9], superconducting circuits [10, 11], and nonlinear photonic systems [12]. Of interest, recent milestone achievements in programmable connectivity in condensed matter using cold atomic gases [13] now permit construction of intriguing networks of coupled elements. However, in general, many lab systems are by their physical nature unable to form unconventional graph topologies. In the past decade, driven-dissipative Bose-Einstein condensates of exciton-polaritons (from here on, polaritons) in planar microcavities have substantially advanced in optical reprogrammability [14–21]. There, each condensate is driven by a focused non-resonant optical excitation beam forming a localized macroscopically coherent wavefunction [22]. The coupling strength between neighbouring condensates is roughly given by their mutual overlap with an exponential fall-off as a function of separation distance [23–25]. This means that nearest-neighbour (NN) coupling dominates over next-nearest-neighbour (NNN) coupling making polariton networks inherently planar in a graph topology sense. Overcoming this spatial coupling hierarchy can offer opportunities to observe spontaneous ordering and emergent polariton effects in non-planar graph topologies [26–32]. However, this is extremely challenging, requiring very fine control over the two-dimensional polariton potential landscape with limitations of its own [33].

In this Letter, we demonstrate that spin-orbit coupled (SOC) exciton-polariton condensates can overcome this challenge. Polaritons are quasiparticles exhibiting intermixed properties of excitons and photons, which appear when light and matter are brought to the strong coupling regime [34]. As a consequence, the photon polarisation is explicitly connected to the polariton pseudospin (or just "spin" for short) with $\hat{\sigma}_z = \pm 1$ spin-projections along the cavity growth axis representing $\sigma^\pm$ circularly polarized light. Their two-component integer spin structure has led to deep exploration into nonequilibrium spinor quantum fluids [35]. Polaritons mostly decay through photons leaking out of the cavity containing all the information on the condensate such as energy, momentum, density, and spin. This salient feature allows direct, yet non-destructive, measurement of the condensate spin distribution using polarization resolved photoluminescence (PL) imaging.

Both the polariton condensate and the incoherent photoexcited background of excitons sustaining it adopt the circular polarisation of the nonresonant excitation [36, 37] due to the optical orientation effect of excitons [38, 39] and spin-preserving stimulated scattering of excitons into the condensate [40]. This permits excitation of a condensate of a well defined macroscopic $S_z \sim \langle \hat{\sigma}_z \rangle$ spin projection [41–44]. Subsequently, the inherent TE-TM splitting of the microcavity [45] will start rotating the spin of any condensate polaritons which obtain finite wavevector and flow away from the pump spot [46, 47]. This is also referred to as the optical spin Hall effect [48, 49]. Namely, the splitting between TE and TM polarized cavity photon modes acts as a directionally dependent in-plane effective magnetic field [48, 50] (i.e., effective SOC [51]) causing the spins of outflowing condensate polaritons to start precessing [see Fig. 1(a) and Fig. 1(b)]. The strength of this effective SOC scales quadratically with the polariton momentum, $\propto k^2$ and can even be electrically tuned [52, 53]. This makes so-called ballistic condensates ideal for enhanced SOC effects [46, 47] due to their extremely high kinetic energies obtained through repulsive Coulomb interactions with the locally pump-induced exciton reservoir. More-over, because of their long-range coherent particle outflow, ballistic condensates can couple over macroscopic distances much greater than their respective full-width-at-half-maximum [24] while also preserving their spin information [44, 46, 47].

Recently, it was theoretically predicted that ballistic condensates could invert their neighbour coupling hierarchy, making NNN stronger than NN, through a spin-screening effect made possible by the effective SOC stemming from TE-TM splitting [54]. Here, we provide experimental evidence of these recent predictions. We present a study of a spinor polariton dyad (two coupled condensates) and a triad [three coupled condensates, see schematic Fig. 1(c)] wherein each condensate ballistically emits a coherent pseudospin current which rapidly precesses as it propagates [46, 47]. We demonstrate control over the coupling strength between neighbouring condensates by changing the spatial distance between them (denoted $d$) relative to the spatial precession period of the condensate pseudospin (denoted $\xi$).

We briefly explain the idea of spin-screened polariton coupling. The three peaks in Fig. 1(c) represent the condensate centers excited by three co-localized Gaussian pump spots of equal intensity. The red-blue colour map shows the precession of the polariton pseudospin as it radially propagates in-plane away from each condensate center, with red representing $S_z = +1$ (spin-up polaritons) and blue representing $S_z = -1$ (spin-down polaritons). The height of the peaks represents the intensity of the condensate emission. The distance between the condensate centers relative to the spatial oscillations of the pseudospin modifies the coupling between them. In the non-screened state [Fig. 1(c)] NN condensates are excited at a distance equal to integer number of periods of pseudospin oscillations, $d = n\xi$ where $n = 1, 2, 3, \ldots$. This means that propagating condensate polaritons arrive at NNs with unchanged spin projection. On the contrary, in the screened state [Fig. 1(d)] NNs are separated by $d = (n-1/2)\xi$ and polaritons arrive at their NNs with opposite spin-projection which reduces the condensate coupling, while coupling between NNNs is still maintained.

The microcavity used in this study consists of a $5\lambda/2$ AlGaAs cavity surrounded by two distributed Bragg mirrors (DBR) of 35 and 32 pairs of AlGaAs/AlAs for the bottom and top DBR correspondingly with the 12 GaAs QWs separated into four sets of three QWs placed at the antinodes of electric field within the cavity. The cavity quality factor is around $Q \sim 16000$ with the corresponding polariton lifetime $\tau_p \approx 5$ ps and Rabi splitting of 9 meV. The measured TE-TM splitting is $\approx 0.2$ meV at $k = 3$ $\mu$m$^{-1}$ in-plane wavevector. Microcavity had the same detuning in all pumping spot positions providing equal values of the condensation threshold for each single isolated condensate. See section S1 in the Supplemental Material [55] for further experimental details.

The normalized Stokes parameters of the cavity emission are written,

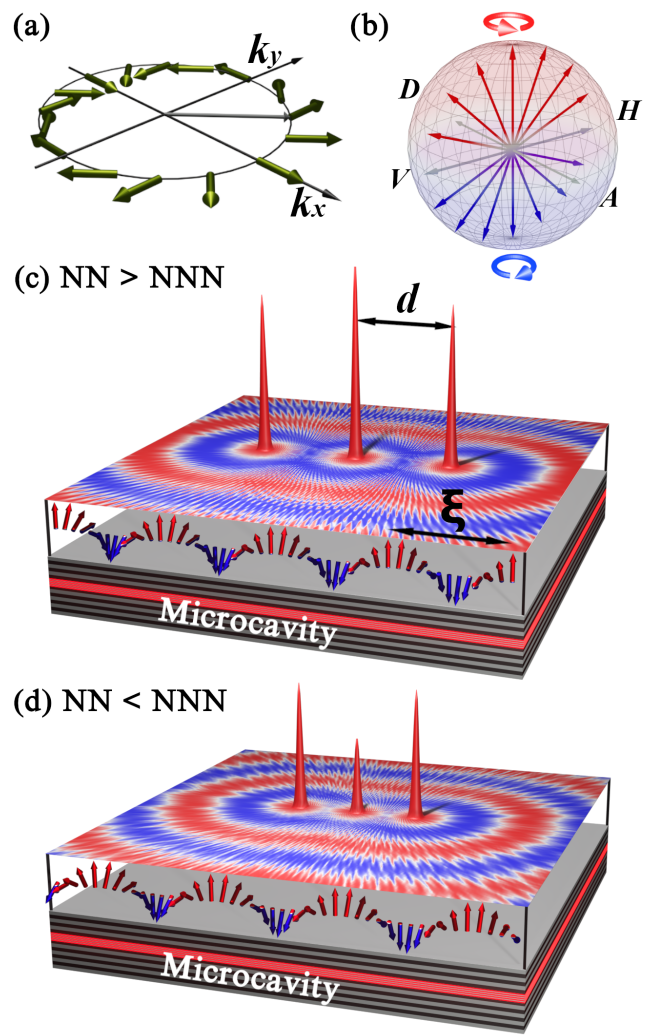

Figure 1. **(a)** Schematic of the effective SOC magnetic field distribution (dark olive arrows) from the TE-TM splitting on a momentum-space circle. **(b)** Schematic of the Poincaré sphere showing example pseudospin precession as polaritons propagate (blue and red arrows). Schematic representing two pump geometries where the distance between the central and edge pump spots equals to **(c)** one full period of pseudospin oscillation (NN is stronger than NNN) and **(d)** half oscillation period (NN is weaker than NNN). The height of the peaks represents the intensity of the condensate emission, and the red, white, and blue colour map shows the precession of the polariton pseudospin propagating in the cavity plane, with red representing $S_z = +1$ (spin-up polaritons) and blue representing $S_z = -1$ (spin-down polaritons). Red and blue arrows show the pseudospin precession of the polaritons propagating from the edge condensates along the triad axis

$$S_{x,y,z}(\mathbf{r}) = \frac{I_{H,D,\sigma^+}(\mathbf{r}) - I_{V,A,\sigma^-}(\mathbf{r})}{I_{H,D,\sigma^+}(\mathbf{r}) + I_{V,A,\sigma^-}(\mathbf{r})}, \quad (1)$$

where $\mathbf{r} = (x, y)$ is the in-plane coordinate and $I_{H(V),D(A),\sigma^+(\sigma^-)}(\mathbf{r})$ corresponds to horizon-

tally(vertically), diagonally(antidiagonally), and right-circularly(left-circularly) polarized (RCP and LCP for short) PL, respectively. Formally, the Stokes parameters relate to the condensate pseudospin through $\mathbf{S} = \langle \Psi^\dagger | \hat{\boldsymbol{\sigma}} | \Psi \rangle / \langle \Psi^\dagger | \Psi \rangle$ where $\Psi = (\psi_+, \psi_-)^\mathrm{T}$ is the condensate spinor order parameter and $\hat{\boldsymbol{\sigma}}$ is the Pauli matrix-vector. The $S_x(\mathbf{r})$ and $S_y(\mathbf{r})$ components represent the degree of linear and diagonal polarisation but are not important in this study (also due to the predominant circular polarisation of the condensates used here). Experimental measurements were reproduced using a generalised two-dimensional Gross-Pitaevskii equation (2DGPE) (see section S2 in the Supplemental Material [55] with references [56, 57] therein).

In Fig. 2 we present results for two polariton condensates separated by $d \approx \xi/2$. Data for a single isolated condensate gives a $S_z$ period around $\xi \approx 90$ $\mu$m (see section S1 in the Supplemental Material [55]). Figures 2(a) and 2(b) show the measured and simulated spatial distribution of the $S_z$ component with spatial pseudospin oscillations clearly visible due to the SOC rotating the spin of the outflowing polaritons. Note that unavoidable dephasing of polaritons in experiment results in lowered $S_z$ values compared to simulations as indicated on the colorbars. Smaller ripple-like modulations are also visible due to the standing wave interference between the two phase-locked condensates as reported before [24, 44, 54]. These ripples are characterized by a small-scale period $\lambda = 2\pi/\langle k_c \rangle \approx 3$ $\mu$m, where $\langle k_c \rangle$ is the average outflow momentum of polaritons from their condensates. In contrast, the large-scale $S_z$ period is given by $\xi = 2\pi/\Delta_k \gg \lambda$ where $\hbar\Delta_k/\sqrt{2\varepsilon_c} = |\sqrt{m_\mathrm{TE}} - \sqrt{m_\mathrm{TM}}|$ and $\varepsilon_c \approx 3$ meV is the condensate energy (measured from $k = 0$ at the dispersion) and $m_\mathrm{TE,TM}$ are the effective masses of TE and TM polarized polaritons [45].

The spin screening effect can be observed as periodic extrema in the integrated PL intensity, which represents the condensate occupation, as a function of separation distance $d$ in Fig. 2(c). At the maxima the coupling is unscreened and NN coupling is strong. At the minima the coupling is screened and NN coupling is weak. Black dots and black solid curve denote experimental measurements and calculations, respectively. In the absence of SOC one would observe monotonically decreasing emission intensity with only short variations (order of $\lambda$) corresponding to in-phase and anti-phase flip-flop transitions between the synchronized condensates [24]. Instead, we observe strong non-monotonic behaviour with clearly visible maxima around 67 and 154 $\mu$m, and minima around 56 and 135 $\mu$m. Notice that the distance between the two maxima and the two minima correlates with the measured $\xi \approx 90$ $\mu$m period of $S_z$ oscillations.

The discrepancy between the absolute locations of the minima and maxima with the predicted critical distances for screened ($\xi/2, 3\xi/2$) and unscreened ($\xi, 2\xi$) coupling, respectively, can be understood as follows. Firstly, when two condensates are coupled their energy is redshifted on average [24] leading to smaller $\varepsilon_c$ and thus larger $\xi$ in the coupled system. Second, the finite width of the pump spots modulates the phase of polaritons and causes a shift in the $S_z$ period. Third, the cavity here has higher levels of disorder than strain-compensated cavities [58] which can affect the spatial coupling. That's why the relative distances between the extrema are more meaningful than their absolute locations. This interpretation is verified in 2DGPE modeling which accurately reproduces the locations of the extrema. Note that the slight discrepancy between modeling and experiment in Fig. 2(c) between 70 and 120 $\mu$m can possibly be attributed to the large parameter space of the 2DGPE making quantitative matching somewhat challenging or localized defects in the sample which scatter outflowing polaritons, consequently decreasing the coupling efficiency. Indeed, the integrated emission intensity of the coupled condensate system is proportional to their non-Hermitian coupling strength from their mutual overlap over the pumped areas, since it determines both the imaginary (and real) part of their complex energies [23]. However, large de-fects are scarce in the sample [46]. In order to verify that the modulation we observe is dominantly coming from the spin screening effect we have carefully chosen a clean part of the sample, with the minimum amount of large defects. We also tested a few other relatively clean sample locations and found qualitatively the same modulation period in the integrated emission dictated by the period of the spin precession.

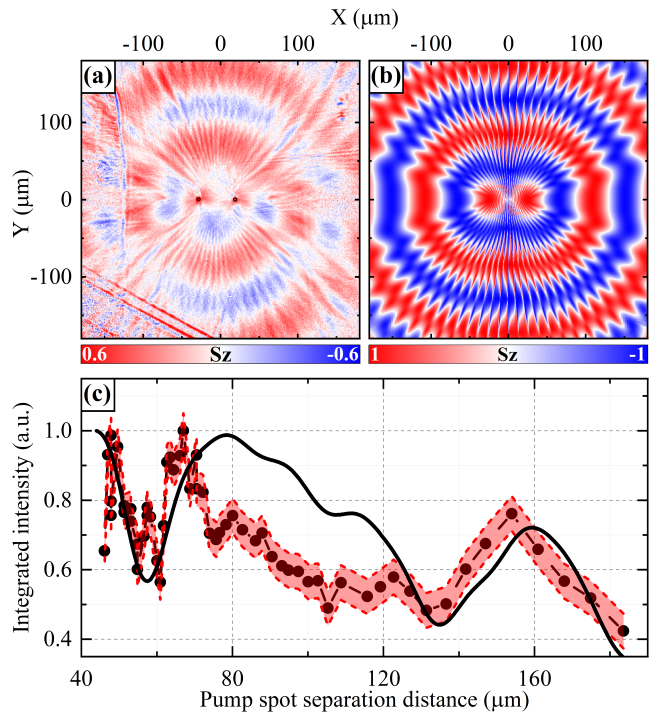

Figure 2. Two polariton condensates. **(a)** Experimentally measured and **(b)** simulated numerically real space $S_z$ component of the Stokes vector of the polariton condensates emission. In panel **(a)** black circles show the position of pump spots. **(c)** Total integrated emission intensity dependence on the separation distance between two condensates pump spots. In panel **(c)** black dots shows the experimentally measured values with red region representing the error of the total intensity value. Black curve shows the same dependence calculated numerically

In order to demonstrate the NNN coupling using the all-optical spin screening effect we investigated the system containing a chain of three condensates similar to the system depicted schematically in Fig. 1. As in the previous experiment with two condensates, all condensates were excited using tightly focused RCP laser pump spots of equal intensity above threshold. Figures 3(a) and 3(b) show the measured and simulated spatial distribution of the three condensate $S_z$ component with NN distance of $d \approx \xi/2$. As in the previous case of two condensates, the system forms a joint macroscopic coherent state resulting in an oscillating $S_z$ pattern elongated along the horizontal axis with three RCP condensate circles of equal degree of polarisation in the centre. Amazingly, the intensity of the central condensate was suppressed relative to the outer ones, evidencing reduced NN coupling due to the spin screening effect, see in Fig. 3(c) measured (red diamonds) and simulated (black solid curve) intensity distribution along the triad axis.

To unambiguously demonstrate the spin screening effect in the triad, we measured (dots) and simulated (solid curve) the dependence of the central condensate intensity as a function of NN separation distance with results presented in Fig. 3(d). Both experiment and calculations show a clear dip around $d = 52$ $\mu$m $\approx \xi/2$, corresponding to spin-screened NN coupling, followed by a small peak around $d = 80$ $\mu$m $\approx \xi$ where the NN coupling is restored. The observed suppression of the central condensate intensity provides strong evidence of spin-screened NN coupling mediated by the spin coherence of the system.

Moreover, we experimentally measured pump power dependence for each separation distance and extracted polariton condensation threshold values, which are shown in Fig. 3(e) (red circles). The horizontal dashed line is the threshold value of the isolated condensate. In the absence of the TE-TM splitting monotonic increase of the threshold value converging to the isolated condensate threshold is expected with the increase of the separation distance between the condensates. In our system we observe maximum threshold at the separation distance of 52 $\mu$m, which precisely corresponded to the minimum of the central condensate intensity in Figs. 3(c) and 3(d). It confirms that the NN condensate interaction is effectively screened at this separation distance due to the TE-TM splitting. Around a separation distance close to the full period of $S_z$ oscillation ($d \approx \xi$) a decrease in the threshold power was observed, as expected with NN coupling restored. A simple linear coupled oscillator model [solid curve in Fig. 3(e)] is able to explain the behaviour of the threshold power (see section S3 in the Supplemental Material [55]).

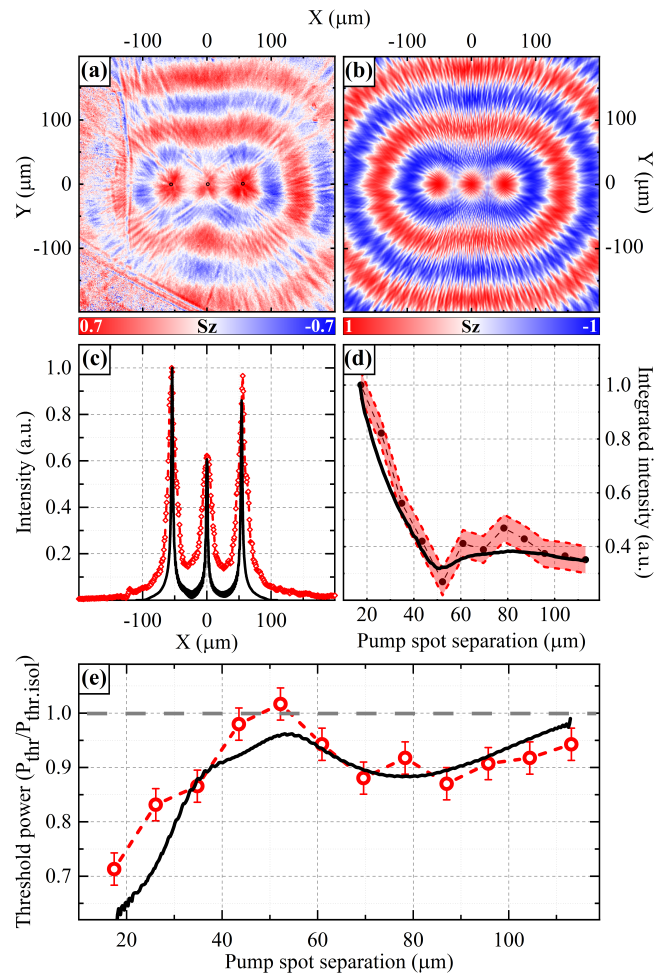

Figure 3. Three polariton condensates. **(a)** Experimentally measured and **(b)** simulated real space $S_z$ component of the Stokes vector of the polariton condensates emission. In panel **(a)** black circles show the position of pump spots. **(c)** Measured experimentally (red diamonds) and calculated numerically (solid black curve) real space intensity distribution along the triad axis. **(d)** Dependence of the central condensate PL intensity on the separation distance between the condensates pump spots measured experimentally (black dots) and calculated numerically (solid black curve); red region represents the error of the total intensity value. The dashed curves are guides to the eye. **(e)** The system threshold power dependence on the separation distance between the condensates pump spots measured experimentally (red circles) and calculated numerically (solid black curve); red bars represent the error. Grey dashed line in panel **(e)** shows the threshold power for single isolated condensate.

In summary, we have experimentally demonstrated that next-nearest-neighbours coupling can be made stronger than nearest-neighbour coupling in ballistically expanding spinor exciton-polariton condensates which was recently proposed in Ref. [54]. This unconventional near-inversion of the spatial coupling hierarchy between condensates stems from the combination of TE-TM splitting and the ballistic polariton flow from each condensate. Outflowing polaritons experience effective spin-orbit coupling which rotates their spin state as they propagate from one condensate to the next. Depending on distance, the overlap (coupling) between the condensates can become spin-screened depending on the polariton spin projection upon arrival at its neighbour. We believe that the demonstrated alteration of the conventional condensate coupling hierarchy could pave the way towards all-optical simulation of many-body ballistic systems belonging to non-planar graph topologies using networks of spinor polariton condensates. In particular, we have noticed a resemblance between the distance-dependent spinor polariton condensate coupling strength and the well known RKKY mechanism (see section S4 in the Supplemental Material [55] with reference [59] therein) which offers perspectives on polaritonic simulation on long-range coupled magnetic dipoles in conductive materials. Another, intriguing effect of making NN polariton coupling strength comparable to NNN couplings in square graphs is the potential to generate stable arrays vortices through XY energy minimization [21] (see section S5 in the Supplemental Material [55] with reference [32] therein).

The authors acknowledge the support of the European Union's Horizon 2020 program, through a FET Open research and innovation action under the grant agreements No. 899141 (PoLLoC) and no. 964770 (TopoLight). H.S. acknowledges the project No. 2022/45/P/ST3/00467 co-funded by the Polish National Science Centre and the European Union Framework Programme for Research and Innovation Horizon 2020 under the Marie Skłodowska-Curie grant agreement No. 945339.

distribution of $S_z$ component of the stokes vector from a single isolated polariton condensate; two dimensional spinor polariton model; theory of the threshold behaviour in a spin screened condensate triad; comparison with RKKY mechanism and discussion on the potential generation of spinor polariton vortical array. The Supplemental Material also contains Refs. [56,57,59].

## S1. EXPERIMENTAL DETAILS, TE-TM SPLITTING AND SINGLE ISOLATED CONDENSATE REAL SPACE $S_z$ COMPONENT OF THE STOKES VECTOR

In this supplemental section we present experimental details, experimentally measured TE-TM splitting and real space distribution of $S_z$ component of Stokes vector of the emission from a single isolated polariton condensate. All measurements were performed at 6 K using a continuous flow cold finger cryostat. We used a right circularly ($\sigma^+$) polarized non-resonant continuous wave laser excitation tuned to the first Bragg minimum of the microcavity reflection spectra at 754 nm. To reduce sample heating we used an acousto-optic modulator driven by rectangular voltage pulse train at 10 kHz repetition rate with 5% duty cycle. A spatial light modulator was used to structure the pump spatial profile into one, two, or three Gaussian spots focused on the sample using a microscope objective lens with 0.4 numerical aperture. In order to obtain simultaneously real space, $k$-space, and spectrally resolved $k$-space images of the time-averaged PL two separate CCD cameras and a 0.75 m monochromator with 1200g/mm diffraction grating equipped with the CCD camera were used. A quarter wave plate and a Wollaston prism were introduced in the optical path for real space imaging to simultaneously measure the right-circularly polarized and left-circularly polarized components of the PL with the same CCD camera.

In Fig. S1(a) we show the $S_z$ spatial oscillations due to the spin-orbit coupling (SOC) rotating the pseudospin of the polaritons propagating away from the condensate excited using single Gaussian spot. The oscillation period $\xi$ was measured to be around 90 $\mu$m with the oscillations amplitude of $\pm 0.6$.

In order to experimentally estimate the value of TE-TM splitting we measured dispersion of the lower polariton branch in a linear regime (i.e., below condensation threshold) along the in-plane $k_\parallel$ momentum axis [see Fig. S1(b)]. Splitting of the dispersion is clearly observed at the higher values of in-plane $k$ vector with the higher energy branch corresponding to the emission from the vertically polarized polaritons. The energy splitting between horizontally and vertically polarized polaritons possesses parabolic dependence on the in-plane momentum and $\approx 0.2$ meV at $k = 3\,\mu\text{m}^{-1}$ in-plane wavevector.

## S2. TWO DIMENSIONAL SPINOR POLARITON MODEL

The experimental observations are reproduced by numerically solving a generalized Gross-Pitaevskii equation (S1) for macroscopic spinor polariton wavefunction $\Psi(\mathbf{r}, t) = (\psi_+, \psi_-)^\text{T}$ coupled to an active exciton reservoir with density $\mathbf{n}_A(\mathbf{r}, t) = (n_{A_+}, n_{A_-})^\text{T}$ rate equation [1],

$$i\frac{\partial \psi_\pm}{\partial t} = \left[-\frac{\hbar \nabla^2}{2m} + \frac{i}{2}\left(R n_{A_\pm} - \gamma\right) + \alpha_1 |\psi_\pm|^2 + \alpha_2 |\psi_\mp|^2 + U_\pm(\mathbf{r}) + V(\mathbf{r})\right]\psi_\pm + \Delta_{LT}\left(\frac{\partial}{\partial_x} \mp i\frac{\partial}{\partial_y}\right)^2 \psi_\mp, \quad \text{(S1)}$$

$$U_\pm = G_1\left(n_{A_\pm} + n_{I_\pm}\right) + G_2\left(n_{A_\mp} + n_{I_\mp}\right), \quad \text{(S2)}$$

$$\frac{\partial n_{A_\pm}}{\partial t} = -\left(\Gamma_A + \Gamma_s + R|\psi_\pm|^2\right)n_{A_\pm} + W n_{I_\pm} + \Gamma_s n_{A_\mp}. \quad \text{(S3)}$$

Here, $\pm$ represents the spin of polaritons and excitons along the cavity growth axis, $m$ is the polariton effective mass in parabolic dispersion approximation, $\gamma$ is the polariton decay rate, $G_1 = 2g|\chi|^2$ and $\alpha_1 = g|\chi|^4$ are the same spin polariton-reservoir and polariton-polariton interaction strengths, respectively, $g$ is the exciton-exciton Coulomb interaction strength, $|\chi|^2$ is the excitonic Hopfield fraction of the polariton, and $\Delta_{LT}$ represents the strength of the TE-TM splitting. Opposite spin interactions, usually much weaker, were chosen to be $G_2 = -0.2 G_1$ and $\alpha_2 = -0.2\alpha_1$

---

* DovzhenkoDS@gmail.com

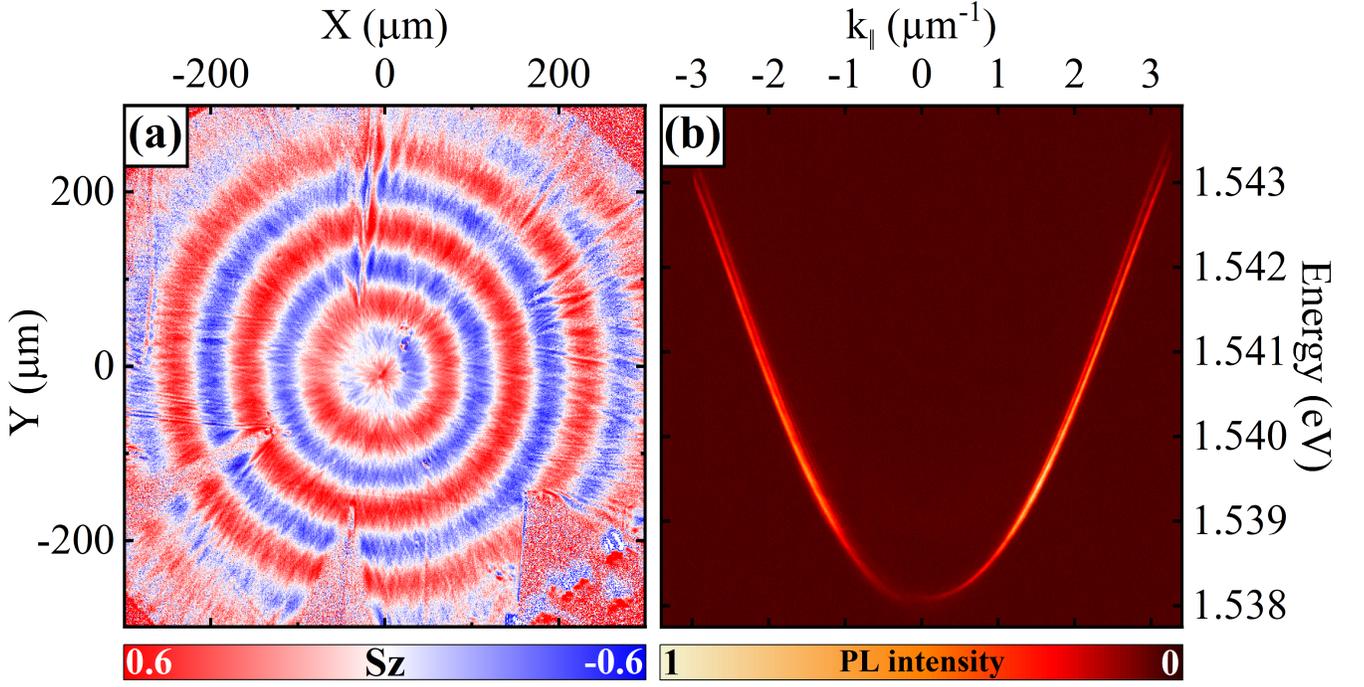

Figure S1. **(a)** Experimentally measured real space $S_z$ component of the Stokes vector of the single isolated polariton condensate emission. **(b)** Energy and in-plane wavevector resolved normalized PL intensity from the lower polariton branch below the threshold

for completeness but we note that our results to not qualitatively depend on these terms. $R$ is the scattering rate of reservoir excitons into the condensate, $\Gamma_A$ is the active reservoir decay rate, and $\Gamma_s$ represents exciton spin relaxation rate [2].

A so-called inactive reservoir of excitons $n_{I,\pm}$ also contributes to the blueshift of polaritons as depicted in Eq. (S2). This reservoir corresponds to high-momentum excitons which do not scatter into the condensate but instead drive the active low-momentum excitons (S3). In continuous wave experiments the inactive reservoir density can be written

$$Wn_{I,+} = \frac{P_0(\mathbf{r})}{W + 2\Gamma_s}(W\cos^2(\theta) + \Gamma_s),$$
$$Wn_{I,-} = \frac{P_0(\mathbf{r})}{W + 2\Gamma_s}(W\sin^2(\theta) + \Gamma_s),$$
(S4)

where $P_0$ is the total power density of the incident coherent light with degree of circular polarization expressed as $S_3 = P_0[\cos^2(\theta) - \sin^2(\theta)] = P_0\cos(2\theta)$. Since our experiment is performed with fully right hand circularly polarized light, we set $\theta = 0$ from here on. The phenomenological parameter $W$ quantifies conversion rate between same-spin inactive and active exciton reservoirs. The pump profile is written as a superposition of Gaussians $P_0(\mathbf{r}) = p_0\sum_n e^{-|\mathbf{r}-\mathbf{r}_n|^2/2w^2}$. To represent tight focusing of excitation beams we used Gaussians with 2 $\mu$m full-width-at-half-maximum.

Lastly, given the disorder present at the large spatial scales of the experiment we include a random potential landscape in our simulation given by $V(\mathbf{r})$ generated as a random Gaussian-correlated potential [3]. The simulation parameters are based on previous GaAs microcavity experiments [4, 5]: $m = 5\times 10^{-5}$ of free electron mass; $\gamma^{-1} = 5.5$ ps; $|\chi|^2 = 0.4$; $\hbar g = 0.5\,\mu\text{eV}\,\mu\text{m}^2$; $R = 3.2g$; $W = \Gamma_A = \gamma$; $\Gamma_s = \gamma/4$; $\Delta_{LT} = 0.036$ ps$^{-1}$ $\mu$m$^2$. The disorder potential was generated with 1.5 $\mu$m correlation length and 0.06 meV root mean squared amplitude.

We note that in order to compensate for additional background noise in experiment (i.e., additional light coming from spontaneous emission of bottleneck excitons) we applied a global shift to the integrated densities of the condensate $|\psi_\pm|^2$ by approximately 10 percent in order to match the experimental values in Figure 3(d) in the main text. This difference between modeling and experiment is more evident in Figure 3(c) where the experimentally measured photoluminescence (PL) intensity is more spread out than simulated condensate densities. This can also come from the finite diffusion of excitons which we have neglected here for simplicity. Nevertheless, the calculated relative amplitude of the PL at the pump positions follows the experimental results quite precisely, which, therefore, justifies the use of current model and provides a clear quantitative evidence of the spin-screening happening in the system.

**S3. THEORY OF THE THRESHOLD BEHAVIOUR IN A SPIN SCREENED CONDENSATE TRIAD**

The behaviour of the pump threshold from experiment in the triad configuration can be reproduced by scrutinizing the eigenenergies of an appropriate linear operator which couples the three condensates together. In other words, we neglect polariton nonlinearities so close to the threshold. The threshold is reached when a single eigenvalue belonging to the three coupled condensates crosses from the lower- to the upper-half of the complex plane.

We will start by defining the state vector of the system,

$$|\Psi\rangle = (\psi_{1,+}, \psi_{1,-}, \psi_{2,+}, \psi_{2,-}, \psi_{3,+}, \psi_{3,-})^{\mathrm{T}}. \tag{S5}$$

Here, the index $n \in \{1, 2, 3\}$ denotes the left, middle, and right condensate, respectively. The spectrum of the coupled system in the linear regime (i.e., close to threshold $|\psi_{n,\pm}|^2 \simeq 0$) can be described with the following non-Hermitian operator separated into three parts for clarity,

$$\hat{H} = \begin{pmatrix} \omega_+ & 0 & 0 & 0 & 0 & 0 \\ 0 & \omega_- & 0 & 0 & 0 & 0 \\ 0 & 0 & \omega_+ & 0 & 0 & 0 \\ 0 & 0 & 0 & \omega_- & 0 & 0 \\ 0 & 0 & 0 & 0 & \omega_+ & 0 \\ 0 & 0 & 0 & 0 & 0 & \omega_- \end{pmatrix} + \begin{pmatrix} 0 & 0 & J_+ & \delta J & 0 & 0 \\ 0 & 0 & \delta J & J_- & 0 & 0 \\ J_+ & \delta J & 0 & 0 & J_+ & \delta J \\ \delta J & J_- & 0 & 0 & \delta J & J_- \\ 0 & 0 & J_+ & \delta J & 0 & 0 \\ 0 & 0 & \delta J & J_- & 0 & 0 \end{pmatrix} + \begin{pmatrix} 0 & 0 & 0 & 0 & K_+ & \delta K \\ 0 & 0 & 0 & 0 & \delta K & K_- \\ 0 & 0 & 0 & 0 & 0 & 0 \\ 0 & 0 & 0 & 0 & 0 & 0 \\ K_+ & \delta K & 0 & 0 & 0 & 0 \\ \delta K & K_- & 0 & 0 & 0 & 0 \end{pmatrix} \tag{S6}$$

The first matrix describes the complex self-energy of each oscillator (condensate) composed of the local pump blueshift ($G$) and gain ($R$), and cavity losses ($\gamma$). This contribution from the pump can be parametrized in terms of the reservoir spin populations,

$$\omega_\pm = \left(G_1 + \frac{iR}{2}\right)(N_{A,\pm} + N_{I,\pm}) - \frac{i\gamma}{2}. \tag{S7}$$

where $\int n_{A(I),\pm}\, d\mathbf{r} = N_{A(I),\pm}$ [i.e., spatially integrating (S3) and (S4)]. Here, we will neglect opposite spin interaction $G_2$ for simplicity.

Each condensate is coupled ballistically with its nearest neighbours with coupling strength $J_\pm$ and next-nearest neighbours with strength $K_\pm$ determined by the overlap between different condensates over their respective pump spots. Approximating the tightly focused pump spots as delta functions, we can write the coupling between the ballistic condensates as [4],

$$J_\pm = \cos(\xi d + \Phi)\left(G_1 + \frac{iR}{2}\right)(N_{A,\pm} + N_{I,\pm})H_0^{(1)}(kd + \phi), \tag{S8}$$

$$K_\pm = \sin(2\xi d + \Phi)\left(G_1 + \frac{iR}{2}\right)(N_{A,\pm} + N_{I,\pm})H_0^{(1)}(2kd + \phi) \tag{S9}$$

The square cosine (sine) modulations in the coupling stem from a pseudospin screening effect coming from the strong influence of TE-TM splitting on the ballistic condensates [6] as explained in the main manuscript. Here, $\xi$ denotes the period of the pseudospin precession for a single condensate in experiment. $H_0^{(1)}(kd)$ is the zeroth order Hankel function of the first kind. The coupling depends on the product $kd$ where $d$ is the separation distance between two pump spots and $k$ is the complex wavevector of the polaritons with mass $m$ propagating outside the pump spot,

$$k \approx k_c + i\frac{\Gamma m}{2\hbar k_c}. \tag{S10}$$

Here, $k_c$ is the average real wavevector of the outflowing polaritons. The finite size of the Gaussian pump spots introduces some lag into the pseudospin precession because outflowing polaritons need to gradually build up momentum as they leave the pump spot. This is captured in the fitting parameter $\Phi$. For the same reason, an overall phase-lag fitting parameter $\phi$ is also needed in the coupling term between the condensates.

The presence of TE-TM splitting also introduces coupling between opposite spin components denoted $\delta J$ and $\delta K$ written in a similar fashion,

$$\delta J = \delta \cos(\xi d + \Phi)\left(G_1 + \frac{iR}{2}\right)(N_{A,+} + N_{I,+} + N_{A,-} + N_{I,-})H_0^{(1)}(kd + \phi), \tag{S11}$$

$$\delta K = \delta \sin(2\xi d + \Phi)\left(G_1 + \frac{iR}{2}\right)(N_{A,+} + N_{I,+} + N_{A,-} + N_{I,-})H_0^{(1)}(2kd + \phi) \tag{S12}$$

Here, $\delta < 1$ is a fitting parameter describing the amount of opposite spin coupling. Diagonalizing $\hat{H}$ for increasing pump power $P_0$ we identify the threshold as the point in which a single eigenenergy crosses from the lower-half to the upper-half of the complex plane. The results are plotted in Fig. 3(e) in the main text (solid curve) alongside the experimental data, normalized in units of threshold power for the single isolated condensates $P_{\text{thr,iso}}$.

## S4. RESEMBLANCE TO RKKY MECHANISM

In the current study, we have presented a path on using spinor polariton networks to simulate many-body phases in condensed matter, with perspectives on exploring novel emergent nonequilibrium phenomena (e.g., spontaneous spin pattern formation) with unconventional coupling hierarchy. This is in a similar spirit to recent works on programmable cold-atom systems [7].

As an example, the long-range ballistic transport of polaritons gives rise to a spin-oscillatory coupling mechanism (as we have demonstrated) reminiscent of the famous RKKY interaction mechanism between magnetic moments in conductive materials such as graphene [8]. To underline this similarity we can write Eq. (S8) as:

$$J_{\pm} \propto \cos\left(\xi d + \Phi\right)|H_0^{(1)}(kd + \phi)|, \tag{S13}$$

where the absolute value takes only the envelope over the fast-oscillating terms in the Hankel function. The RKKY interaction mechanism is written:

$$J_{\text{RKKY}} \propto \frac{x \cos(x) - \sin(x)}{x^4} \tag{S14}$$

where $x = \xi d$. In Fig. S2 we compare the similarities between Eq. (S8) and Eq. (S14) which underlines the potential of using ballistic polaritons to explore long-range spin phenomena, but now in an optical setting supplemented with polariton nonlinearities.

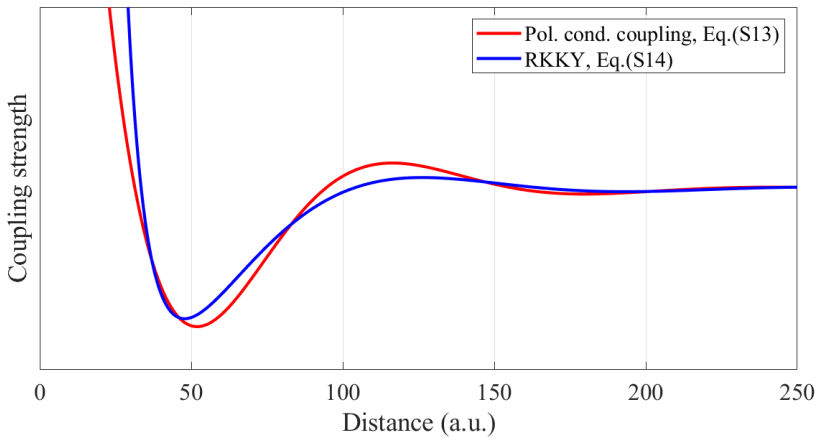

Figure S2. Comparison of the spinor condensate coupling strength and the RKKY mechanism as a function of distance.

## S5. POTENTIAL GENERATION OF SPINOR VORTICAL POLARITON ARRAYS

It is well established that interacting polariton condensates synchronize into phase patterns that correlated with the low energy solution of the XY spin Hamiltonian [9]. In this sense the phasor of a scalar condensate represents a planar spin. This implies that polariton condensates can form a heuristic optimization tool facilitated by the ultrafast dissipative dynamics of polaritons. Recently we showed that projection algorithms (i.e., so-called semi-definite programming techniques) could be used on the polariton condensate phasors ("XY spins") to heuristically solve the max-3-cut problem (and possibly the max-2-cut problem) [10]. The latter is mappable to the problem of finding the ground state of an Ising spin glass, which is a ubiquitous problem associated with the traveling salesman problem, knapsack problem, graph partitioning, etc.

Unfortunately, we do not currently have the experimental scalability to solve practical hard mathematical problems. But there are still interesting ground state solutions of the XY model that can be addressed by our spinor system in the near future. Consider e.g. the square graph shown in Fig. S3(a) where each black node represents a condensate, a red spin its phasor of the dominant spin component, and each edge (line) represents a coupling. If NN couplings are dominant and positive ($J > 0$) then the ground state of the XY Hamiltonian

$$H_{\text{XY}} = -\sum_{n,m} J_{n,m} \cos(\theta_n - \theta_m) \tag{S15}$$

is simply $\theta_n = \theta_m$ for all $n, m$. This is the ferromagnetic state as shown by the red arrows. If, on the other hand, the NNN couplings are made comparable to the NN and have opposite sign ($J < 0$) then the ground state corresponds to a vortex solution where the spin (phasor) rotates around the square [see Fig. S3(b)]. We have performed 2DGPE simulation using parameters of our system and verified that indeed these vortex solutions exist in a stable form in the dominant spin component of the condensate when the NNN couplings becomes strong [see Fig. S3(c) and S3(d)]. This finding is quite interesting since it potentially offers a solution to create large-scale polariton vortex arrays in large square pumping lattices [11] using our techniques.

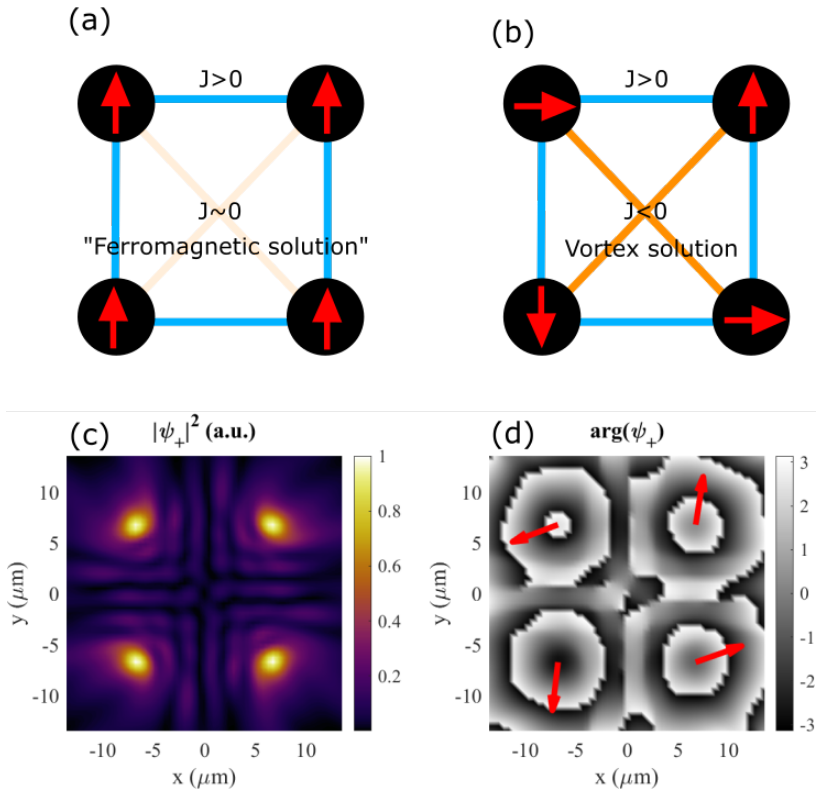

Figure S3. Nontrivial ground state solution in a square graph with NNN couplings. Top panels show schematically the coupling configurations with red arrows depicting the macroscopic phasors ("XY spins") of each condensate. Bottom panels show example condensate density and phase from a 2DGPE simulation producing a vortex state by making NN and NNN couplings comparable.